\documentclass[conference]{IEEEtran}
\IEEEoverridecommandlockouts
\usepackage{cite}
\usepackage{amsmath,amssymb,amsfonts}
\usepackage{algorithmic}
\usepackage{graphicx}
\usepackage{textcomp}
\usepackage{xcolor}
\usepackage{orcidlink}
\usepackage{multicol}
\usepackage{multirow}
\usepackage{float}
\usepackage{subcaption}
\usepackage{listings}
\def\BibTeX{{\rm B\kern-.05em{\sc i\kern-.025em b}\kern-.08em
    T\kern-.1667em\lower.7ex\hbox{E}\kern-.125emX}}
\begin{document}

\title{UNION: A Lightweight Target Representation \\for Efficient Zero-Shot Image-Guided Retrieval\\with Optional Textual Queries}

\author{\IEEEauthorblockN{Hoang-Bao Le$^1$, Allie Tran$^1$, Binh T. Nguyen$^2$, Liting Zhou$^1$, Cathal Gurrin$^1$}
\IEEEauthorblockA{$^1$\textit{ADAPT Centre, School of Computing, Dublin City University} \\
$^2$\textit{Ho Chi Minh University of Science, Vietnam National University} \\
Dublin, Ireland \\
bao.le2@mail.dcu.ie, \{liting.zhou, cathal.gurrin\}@dcu.ie}
}
% \and

\maketitle

\begin{abstract}
    Image-Guided Retrieval with Optional Text (IGROT) is a general retrieval setting where a query consists of an anchor image, with or without accompanying text, aiming to retrieve semantically relevant target images. This formulation unifies two major tasks: Composed Image Retrieval (CIR) and Sketch-Based Image Retrieval (SBIR). In this work, we address IGROT under low-data supervision by introducing UNION, a lightweight and generalizable target representation that fuses the image embedding with a null-text prompt. Unlike traditional approaches that rely on fixed target features, UNION enhances semantic alignment with multimodal queries while requiring no architectural modifications to pretrained vision-language models. With only 5,000 training samples—from LlavaSCo for CIR and Training-Sketchy for SBIR—our method achieves competitive results across benchmarks, including CIRCO mAP@50 of 38.5 and Sketchy mAP@200 of 82.7, surpassing many heavily supervised baselines. This demonstrates the robustness and efficiency of UNION in bridging vision and language across diverse query types. Our dataset and code are available at \href{https://github.com/baohl00/UNION_for_IGROT}{https://github.com/baohl00/UNION\_for\_IGROT}.
\end{abstract} 


\maketitle

\section{Introduction}

Composed Image Retrieval (CIR) and Sketch-Based Image Retrieval (SBIR) are two key tasks in content-based image retrieval that involve retrieving a target image based on an input query. In CIR, the query consists of a reference image and a textual modification describing how the target should differ~\cite{guo2019fashion,Liu2021cirr,baldrati2023zero}, while SBIR uses a sketch to retrieve a matching natural image~\cite{yelamarthi2018zssbir,zhang2016tuberlin}. Although these tasks appear different, they share a common structure: an image-driven query with optional auxiliary language input. We unify them under the general formulation of \textbf{Image-Guided Retrieval with Optional Text (IGROT)}, where the query always contains an anchor image and may be optionally augmented with descriptive text. IGROT reflects a practical retrieval setting that accommodates both visual-only and vision-language queries within a shared framework.

Despite substantial progress in CIR, developing large-scale, high-quality datasets remains challenging. Creating CIR benchmarks requires carefully curated triplets of reference image, modification text, and target image, which is labor-intensive and often domain-specific. As a result, most datasets suffer from limited coverage, ambiguous language, and lack of diversity in compositional relations. To reduce this reliance on manual annotation, some recent works adopt zero-shot or weakly supervised paradigms~\cite{Vo_2019_CVPR,zhang2024magiclens,gu2024compodiff,huynh2025collm}, often using millions of synthetic or web-mined triplets. However, this introduces its own trade-offs: massive data generation is computationally expensive and may contain noisy or unnatural descriptions.

To address these limitations, we construct \textbf{LlavaSCo}, a compact and high-quality dataset derived from LaSCo~\cite{Levy2024lasco}, consisting of approximately 5,000 triplets enhanced with detailed captions generated by the vision-language model LLaVA~\cite{liu2023improvedllava}. This small-scale dataset enables effective CIR training under limited supervision while maintaining semantic richness. Complementarily, for the SBIR setting, we create the \textbf{Training-Sketchy} dataset from Sketchy \cite{liu2017sketchy}, which consists of 5,000 sketch-natural image pairs covering diverse object categories, allowing us to validate our method across both language-augmented and visual-only retrieval scenarios.

While prior CIR research has primarily focused on improving the fusion of the query image and modification text to produce a rich compositional embedding, comparatively little attention has been given to how the target image is represented—which in most works is simply the visual feature extracted directly from a pretrained vision-language model, such as CLIP or BLIP, without further adaptation. Most recent methods~\cite{liu2023zeroshot,gu2024lincir,Levy2024lasco,huynh2025collm} compare the query against fixed target features directly extracted from pretrained vision-language encoders such as CLIP~\cite{radford2021learning} or BLIP~\cite{li2022blip}. However, these fixed embeddings may not be well-aligned with the compositional semantics of the query, especially in fine-grained or relational changes.

To address this, we introduce \textbf{UNION}, a simple yet effective target feature representation. Instead of comparing queries to frozen image embeddings, UNION computes a composed target feature by combining the visual representation of the target image with a null-text prompt, passed through a lightweight Transformer and MLP. This design allows the target representation to reside in the same vision-language embedding space as the query, enhancing semantic alignment and improving contrastive training dynamics. Importantly, UNION does not require modifications to the backbone encoder and supports image-only and image-text retrieval.

\textit{In this paper, our contributions are summarised as follows:}
\begin{itemize}
    \item We demonstrate that strong CIR performance can be achieved using only 5,000 high-quality triplets from LaSCo, refined into our LlavaSCo dataset with detailed captions generated by LLaVA, significantly reducing the dependence on large-scale annotations.
    \item We show that our method generalises to the SBIR setting by treating sketches as image-only queries and applying the same architecture, achieving strong performance with minimal adaptation.
    \item We propose \textbf{UNION}, a novel and efficient target feature representation that combines visual and null-text inputs, improving semantic compatibility with fused queries without requiring changes to the pretrained vision-language backbone.
\end{itemize} 

\section{Related Work}

\subsection{Zero-Shot Image-Guided Retrieval with Optional Text (IGROT)}

The task of Image-Guided Retrieval with Optional Text (IGROT) unifies Composed Image Retrieval (CIR) and Sketch-Based Image Retrieval (SBIR) by allowing queries that contain an anchor image with or without accompanying text. While supervised CIR and SBIR methods often require large-scale annotated triplets, recent efforts aim to reduce supervision via zero-shot or weakly supervised approaches.

\paragraph{Composed Image Retrieval (CIR)}  
To mitigate the cost of manual annotations in CIR, several works have proposed training strategies that eliminate the need for annotated triplets. Pic2Word~\cite{saito2023pic2word} introduced the Zero-Shot CIR (ZS-CIR) setting by learning from weakly labeled image-caption pairs and unlabeled image collections. Other methods focus on scaling data through synthetic triplet generation. For example, MagicLens~\cite{zhang2024magiclens} created 36.7M triplets from web images and PaLM2-generated captions~\cite{anil2023palm}, while CC-CoIR~\cite{ventura2024covr} constructed 3.3M triplets from Conceptual Captions~\cite{sharma2018conceptual}. CoLLM~\cite{huynh2025collm} leveraged large language models (LLMs) like Claude 3 Sonnet\footnote{\url{https://www.anthropic.com/claude/sonnet}} to generate high-quality relational captions for 3.4M image pairs. In contrast, LaSCo~\cite{Levy2024lasco} proposed a lightweight method using GPT-3~\cite{brown2020gpt} and data roaming to collect 360k image-text triplets. Recently, LoGra-Med~\cite{nguyen2024logramedlongcontextmultigraph} reduced the dependence on large-scale datasets by constructing a smaller but more informative training set, using only 10\% of the original data while preserving performance. Following this, our work departs from these scale-intensive strategies by using only 5,000 triplets from LaSCo. We enhance this subset with detailed image captions generated by LLaVA~\cite{liu2023improvedllava}, resulting in a compact yet semantically rich dataset we name \textbf{LlavaSCo}.

\paragraph{Sketch-Based Image Retrieval (SBIR)}  
SBIR has also seen interest in zero-shot generalisation. Earlier methods such as SAKE~\cite{liu2019semantic} focused on preserving discriminative semantic features across modalities. Hybrid fusion~\cite{wu2025hybridInfoFusion} and adapter-based structural alignment~\cite{zhang2025adapter} further reduced domain gaps. DCDL~\cite{li2025dcdl} used causal disentanglement to improve cross-domain representation, and ZSE-SBIR~\cite{lin2023zsesbir} aligned sketches and photos via local patch correspondences for explainable matching. However, these methods rely exclusively on visual cues, which can limit the model’s ability to distinguish subtle semantic differences or relational concepts—particularly when the query includes nuanced modifications that are better captured through language. In contrast, we introduce \textbf{language as an auxiliary modality} in SBIR by pairing sketch queries with a fixed textual prompt during training. This allows us to unify CIR and SBIR under the IGROT framework, using the same model and training strategy to support both query types.

\subsection{Target Representation in Vision-Language Retrieval}

While much of the zero-shot CIR literature focuses on improving query fusion (\cite{liu2023zeroshot}, \cite{gu2024lincir}),  or expanding training data (\cite{huynh2025collm}, \cite{byun2024reducingtaskdiscrepancytext}, \cite{Levy2024lasco}), the target representation has received relatively little attention. Most approaches compare the composed query against fixed target embeddings obtained directly from vision-language models such as CLIP~\cite{radford2021learning} or BLIP~\cite{li2022blip}. These static features may not align well with the compositional semantics of the query, especially in fine-grained or cross-modal scenarios. To address this, we propose the \textbf{UNION} feature, which fuses the target image embedding with a null-text prompt using a lightweight Transformer-MLP stack. This produces a semantically enriched representation that resides in the same embedding space as the composed query, improving retrieval quality without modifying the backbone model or requiring additional supervision.

\section{Methodology}

\begin{figure}[!h]
    \centering
    \includegraphics[width=\linewidth]{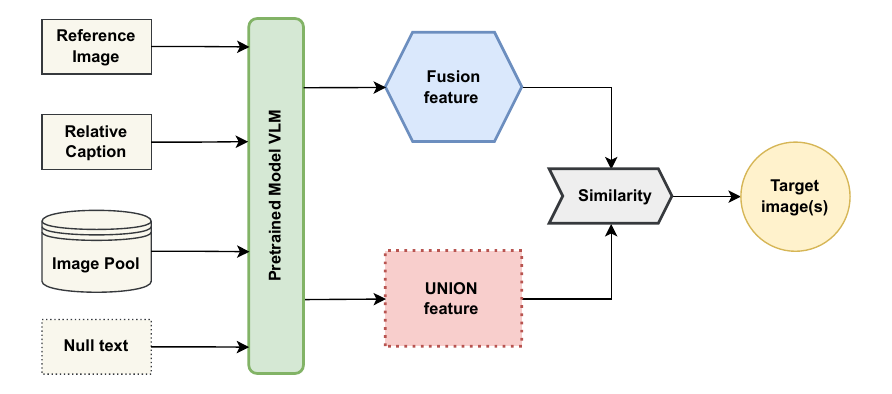}
    \caption{An Overview of Model Architecture in Composed Image Retrieval. Normally, the models do not have Null Text and UNION Feature and directly rank the similarity score of Fusion feature with Image Pool's embedded feature.}
    \label{fig:enter-label}
\end{figure}

In this section, first, we present TransAgg \cite{liu2023zeroshot} as a representative example, to provide an overview of the ZS-CIR paradigm. Following this, we present a detailed explanation of the proposed UNION feature combining the pool's image and null text. For the ZS-CIR task, we introduce \textbf{LlavaSCo}, an updated version of LaSCo~\cite{Levy2024lasco}, which improves both training quality and performance on standard benchmarks. For the ZS-SBIR task, we construct \textbf{Training-Sketchy}, a compact training set derived from Sketchy~\cite{liu2017sketchy}, containing 5,000 high-quality samples.

\subsection{Overview}

In Composed Image Retrieval (CIR), the goal is to retrieve one or more target images \( T \) that satisfy a user query composed of a reference image \( R \) and a textual description \( C \) specifying the desired transformation. A common modeling approach formulates this task as learning a fusion function \( \mathcal{F}_{rc}\in \mathbb{R}^{B\times D} \) with $r\in R$ and $c\in C$, which maps the image-text pair into a joint embedding space. The resulting query representation is then compared against the embedded target image feature \( \textbf{e}_{t_i} \in \mathbb{R}^{B\times D} \), typically using cosine similarity.

To train this representation, we adopt the Batch-Based Classification (BBC) loss~\cite{Vo_2019_CVPR}, which encourages each fused query \( \mathcal{F}_{rc} \) to be most similar to its corresponding target \( \textbf{e}_{t} \) and dissimilar to other targets in the batch. Formally, for a batch of size \( B \), the loss is defined as:

\begin{align}\label{loss}
    \mathcal{L} = -\dfrac{1}{B}\sum_{i=1}^{B} \log \left[ 
        \dfrac{\exp\left(\operatorname{sim}(\mathcal{F}_{r_ic_i}, \textbf{e}_{t_i})/\tau\right)}
              {\sum_{j=1}^{B} \exp\left(\operatorname{sim}(\mathcal{F}_{r_ic_i}, \textbf{e}_{t_j})/\tau\right)}
    \right],
\end{align}

where \( \operatorname{sim}(a, b) = \dfrac{a^\top b}{\|a\| \|b\|} \) denotes cosine similarity, and \( \tau \) is a temperature scaling parameter.

\subsection{UNION: Unified Target Representation}

In standard retrieval setups, models typically compare a multimodal query embedding with a fixed image-only target embedding \( \mathbf{e}_{t_i} \). However, in Composed Image Retrieval (CIR), the query consists of a joint representation of image and text, leading to a modality mismatch when compared to a unimodal target feature. This asymmetry makes it difficult for the model to learn fine-grained semantic alignments, especially when the modification involves subtle relational or attribute-level changes. Moreover, the lack of textual conditioning in target features hinders the model's ability to reason about abstract or compositional semantics.

To bridge this gap, we propose the \textbf{UNION} feature—a unified and semantically enriched target representation that incorporates both the visual and latent textual context. Specifically, we concatenate the target image embedding \( \mathbf{e}_{t_i} \) with a null-text embedding \( \mathbf{e}_\eta \), where \( \eta \) is an empty string passed through a pretrained vision-language model (e.g., CLIP or BLIP). The null-text embedding introduces a neutral linguistic prior that implicitly brings the target closer to the multimodal space used for queries, without injecting external textual content.

\begin{figure}[!h]
    \centering
    \includegraphics[width=\linewidth]{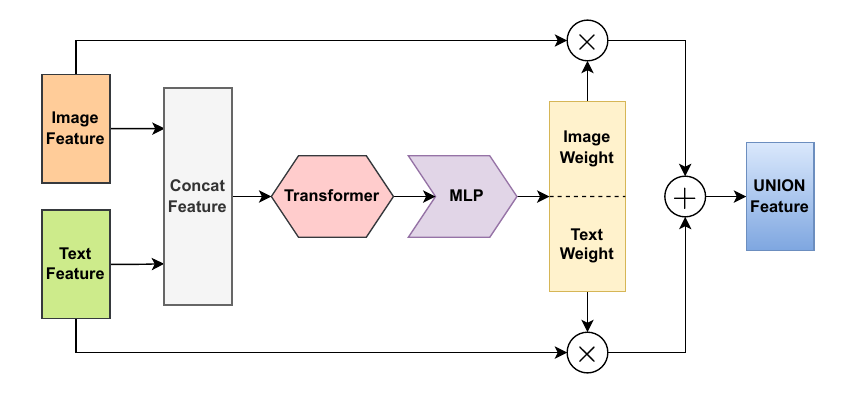}
    \caption{UNION architecture combining all images to be retrieved and Null Text.}
    \label{fig:union}
\end{figure}

As shown in Figure~\ref{fig:union}, the concatenated feature 
\[
\mathbf{e}_{t\eta} = [\mathbf{e}_{t}; \mathbf{e}_\eta] \in \mathbb{R}^{B\times 2\times D},
\]
where \( B \) is the batch size and \( D \) is the feature dimension, is passed through a lightweight Transformer encoder (following the architecture of T5~\cite{radford2021t5}, similar to UNIIR~\cite{wei2024uniir}). The Transformer processes the sequence and outputs a tensor of the same shape \( \mathbb{R}^{B \times 2 \times D} \), which is then pooled (e.g., using average or the first token) to produce a single feature vector per instance \( f^{t\eta} \in \mathbb{R}^{B \times D} \). This vector is passed through a Multi-Layer Perceptron (MLP) to compute the adaptive weight:

\begin{align}
    f^{t\eta} &= \texttt{Transformer}(\mathbf{e}_{t\eta}) \in \mathbb{R}^{B \times 2 \times D} \\
    w_{t} &= \texttt{MLP}(\texttt{Pool}(f^{t\eta})) \in \mathbb{R}^{B \times D}, \quad w_\eta = 1 - w_{t}
\end{align}

% These weights are applied element-wise to the original features \( \mathbf{e}_{t} \) and \( \mathbf{e}_\eta \) to construct the final UNION representation.

The final UNION feature is a learned interpolation between the image and null-text features:

\[
\mathcal{U} = w_{t} \cdot \mathbf{e}_{t} + w_\eta \cdot \mathbf{e}_\eta
\]

% Unlike static alpha blending, this adaptive weighting allows the model to determine how much semantic contribution to draw from visual versus null-text context—enabling flexible fusion across tasks. The Transformer and MLP used here are lightweight (2–3 layers each) and trained jointly with the rest of the model, while all pretrained vision-language backbones remain frozen.

During inference, UNION continues to use the same null-text conditioning, which does not require any additional user input or textual description, ensuring consistent behavior across CIR and SBIR, including sketch-only queries. This structure improves retrieval quality by aligning the target embeddings more closely with the fused query representations and supports plug-and-play integration with various pre-trained backbones.

To train the model, we simply replace the embedding of the raw target image \( \mathbf{e}_{t_i} \) in the loss function (Equation~\ref{loss}) with the UNION feature \( \mathcal{U}_i \), resulting in the updated objective:

\begin{align}\label{loss_new}
    \mathcal{L} = -\dfrac{1}{B}\sum_{i=1}^{B} \log \left[ 
        \dfrac{\exp\left(\operatorname{sim}(\mathcal{F}_{r_ic_i}, \mathcal{U}_i)/\tau\right)}
              {\sum_{j=1}^{B} \exp\left(\operatorname{sim}(\mathcal{F}_{r_ic_i}, \mathcal{U}_j)/\tau\right)}
    \right]
\end{align}

% This UNION-based formulation leads to better semantic grounding of targets, while maintaining compatibility with low-data regimes and frozen vision-language backbones.

\subsection{Dataset Construction}

\subsubsection{LlavaSCo: Caption Refinement for Stronger Alignment.}  
While LaSCo~\cite{Levy2024lasco} provides a large-scale dataset for CIR, we observe that its relative captions often lack specificity or semantic clarity, weakening the learning signal for tasks requiring precise compositional understanding. This is particularly problematic for our UNION-based target representation, which benefits from high-quality text-to-image alignment.

\begin{figure}[!ht]
    \centering
    \includegraphics[width=\linewidth]{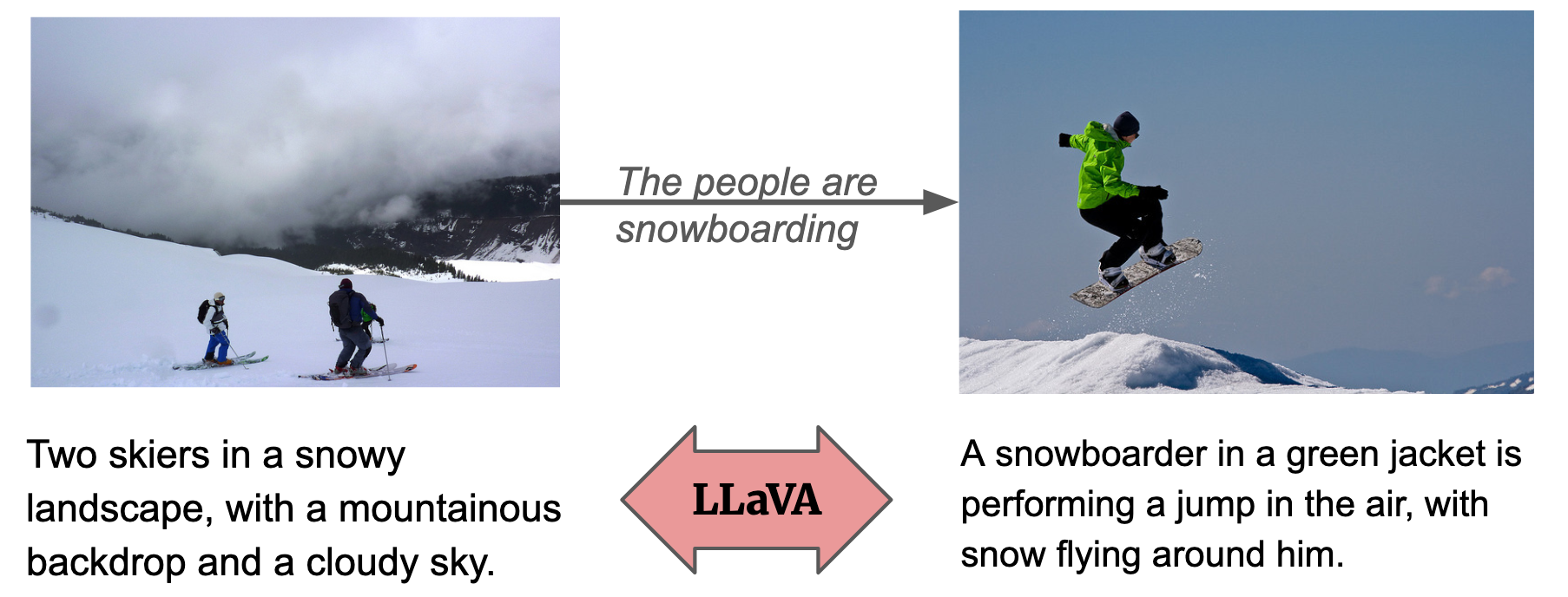}
    \caption{LLaVA Caption for the target image and also the reference image.}
    \label{fig:llava}
\end{figure}

To mitigate this, we enhance a subset of LaSCo using LLaVA~\cite{liu2023improvedllava}, a vision-language model capable of generating detailed captions. For each triplet, we generate a refined target image caption and append it to the original relational caption to produce a more informative instruction. We refer to this enhanced subset (5,000 training triplets) as \textbf{LlavaSCo}, which serves as the training data for our CIR experiments. An example is shown in Figure~\ref{fig:llava}. With the generated caption, the instruction that shifts the image input into the retrieved image is less ambiguous and well-described.

Additional details and examples of the caption generation process are included in the Appendix~\ref{appendix_a:llavasco}.

% From $~360k$ triplets of LaSCo \cite{Levy2024lasco}, we apply LLaVA \cite{liu2023improvedllava} to generate the description for the target images. To construct \textbf{LlavaSCo}, we choose $5,000$ triplets from the training set and maintain the amount of the testing set. 

\subsubsection{Training-Sketchy: Constructing SBIR Triplets}

As a case study within the IGROT framework, we focus on Sketch-Based Image Retrieval (SBIR), where the query is a sketch and the goal is to retrieve a corresponding real-world image. Although prior ZS-SBIR work often constructs separate training sets for each benchmark (e.g., Sketchy, TU-Berlin, and QuickDraw), our approach uses a unified training set. Specifically, we use the Sketchy dataset~\cite{liu2017sketchy}, which provides sketch-image pairs across 125 categories, and we build a small, curated dataset—Training-Sketchy—from the Sketchy dataset only, and apply it across all Zs-ZS-SBIR benchmarks. 
We randomly select 50 categories from the official training split and sample 100 sketch-photo pairs per category, resulting in a total of 5,000 training triplets. To align with our vision-language architecture, we associate each sketch query with a fixed instruction: 

\begin{quote}
    \centering
    \textit{``a real image of sketch''}
\end{quote}

This simple prompt allows sketches to be treated similarly to CIR queries that combine an image with text, enabling cross-task training under the IGROT setting. We refer to this dataset as \textbf{Training-Sketchy}, which we use to train our model on SBIR alongside CIR data from LlavaSCo.

\section{Experiment and Discussion} 

\subsection{Evaluation Datasets and Metrics}

\paragraph{\textbf{Zero-Shot Composed Image Retrieval}}

In the training stage, we use the LlavaSCo dataset—our enhanced version of LaSCo~\cite{Levy2024lasco}—which shares the same index set. We benchmark our method on three widely-used CIR datasets: FashionIQ\cite{guo2019fashion}, which includes over 2,005 triplets across three fashion categories (Dress, Shirt, and Toptee) and a retrieval pool of 5,179 images; CIRR\cite{Liu2021cirr}, comprising 4,148 image-caption queries and 2,316 target images; and CIRCO\cite{baldrati2023zero}, a large-scale benchmark with 800 queries and 123,403 candidate images. Following prior work\cite{zhang2024magiclens}, we evaluate retrieval performance using Recall@K for FashionIQ and CIRR, and mean Average Precision (mAP) for CIRCO. Crucially, during both training and inference, our approach leverages a null-text prompt when constructing target features using the UNION representation. 
Based on prior work \cite{zhang2024magiclens}, we adopt Recall at $K$ ($R@K$) for FashionIQ and CIRR, and mean Average Precision (mAP) for CIRCO as the evaluation metric. 

\paragraph{\textbf{Zero-Shot Sketch-Based Image Retrieval}}

We train the model with the fixed instruction for each sketch query and utilise null text $c=``"$ for inference stage. For testing, we evaluate the model on three benchmarks: Sketchy \cite{yelamarthi2018zssbir} has 12,694 queries dividing into 21 classes from ImageNet-1k and a 12,694-image pool; TUBerlin \cite{zhang2016tuberlin} consists of 30 categories with 2,400 sketches and a corresponding index set has 27,989 images;  QuickDraw \cite{liu2017sketchy} includes 92,291 queries covering 30 classes and a 54,146-sized index set. Following prior work \cite{li2025dcdl}, we report mAP and Precision at $K$ (Prec@K) for each dataset. 

To demonstrate the efficiency of the UNION feature, we also consider the other metrics in Appendix \ref{appendix_b:metrics}.

\begin{table*}[!ht]
    \centering
    \begin{tabular}{|c|c|c|c|c|c|c|c|c|c|}
        \hline \multirow{2}{*}{\textbf{Method}}& \multirow{2}{*}{\textbf{Backbone}} & \multirow{2}{*}{\textbf{\# Params}} & \multirow{2}{*}{\textbf{\# Triplets}} & \multicolumn{2}{|c|}{\textbf{FashionIQ (R)}} & \multicolumn{2}{|c|}{\textbf{CIRR (R)}} & \multicolumn{2}{|c|}{\textbf{CIRCO (mAP)}} \\
        \cline{5-10} & & & & \textbf{@10} &  \textbf{@50} & \textbf{@10} & \textbf{@50} & \textbf{@10} & \textbf{@50} \\
        \hline Pic2Word \cite{saito2023pic2word} & CLIP-L & 429M & 3M & 24.7 & 43.7 & 65.3 & 87.8 & 9.5 & 11.3 \\
        \hline i-SEARLE \cite{agnolucci2024isearle} & CLIP-L & 442M & 205K & 29.2 & 49.5 & 66.7 & 88.8 & 13.6 & 16.3\\
        \hline CIReVL \cite{karthik2024cirevl} & CLIP-L & 12.5B & - & 28.6 & 48.6 & 64.9 & 86.3 & 19.1 & 20.9\\
        \hline MLLM-I2W \cite{bao-etal-2025-mllm} & CLIP-L & - & 3M & 30.3 & 50.1 & 68.4 & 92.4 & - & -\\
        \hline PLI \cite{chen2023pretrainlikeinferencemasked} & CLIP-L & 428M & 695K & \textcolor{red}{35.4} & \textcolor{red}{57.4} & 69.3 & 89.8 & 14.2 & 16.4\\
        \hline LinCIR \cite{gu2024lincir} & CLIP-L & - & 5.5M & 26.4 & 46.6 & 66.9 & 88.8 & 13.9 & 16.2 \\
        \hline MagicLens\cite{zhang2024magiclens} & CLIP-L & 465M & 36.5M & 30.7 & 52.5 & 74.4 & 92.6 & 30.8 & 34.4 \\
        \hline CoLLM \cite{huynh2025collm} & BLIP-B & - & 3.4M & \textcolor{blue}{34.6} & \textcolor{blue}{56.0} & \textcolor{red}{78.6} & \textcolor{red}{94.2} & 20.4 & 23.1 \\
        \hline TransAgg \cite{liu2023zeroshot} & BLIP-B & 235M & 32K & 34.4 & 55.1 & \textcolor{blue}{77.9} & \textcolor{blue}{93.4} & \textcolor{blue}{32.2} & \textcolor{blue}{36.2} \\
        \hline TransAgg + UNION & BLIP-B & 235M & 5K & 31.9 & 51.5 & 77.6 & 92.9 & \textcolor{red}{34.5} & \textcolor{red}{38.5} \\
        \hline
    \end{tabular}
    \caption{Comparison of our method against baseline on three benchmarks of ZS-CIR task. While we reproduce the results of TransAgg on CIRCO, the others are from the original papers. \textcolor{red}{Red} and \textcolor{blue}{Blue} numbers indicate the best and second-best results.}
    \label{tab:cir_overall}
\end{table*}

\subsection{Implementation Details}

Our framework is implemented with Pytorch. We follow the TransAgg setups \cite{liu2023zeroshot} and use the transformer-based 2 layer fusion module with 8 heads and GELU activation. For visual and text encoders, we utilise three popular pretrained models as OpenAI CLIP-B/32 and CLIP-L/14 \cite{radford2021learning}, and BLIP with ViT-B \cite{li2022blip}. In the UNION architecture, we use a 2-layer transformer architecture similar to the T5 Transformer \cite{radford2021t5} with 2 layers, but 8 attention heads for CLIP$_{base}$ and BLIP, and 12 attention heads for CLIP$_{large}$ with each head having $64$ dimensions. In the loss function, we set $\tau = 0.01$ similar to TransAgg \cite{liu2023zeroshot} settings. For a fair comparison, all experiments use $224\times 224$ images to ensure, and additionally, we only choose the results used the aforementioned pre-trained models. To compare with our UNION feature $\mathcal{U}$, we consider the original target image's feature $\textbf{e}_{t}$ and the sum feature $\mathbf{e}_{t\eta}$ of target image $t$ and null text $\eta$. 

The model is optimised with AdamW \cite{loshchilov2017adamw} optimiser with a weight decay of $1e^{-2}$. All experiments are conducted with 2 epochs using learning rate $1e^{-4}$ and a batch size of $32$ on one NVIDIA A100 80GB. For the generated captions in LlavaSCo, we adapt the vision language model LLaVA-v1.6-Mistral-7B. 

\subsection{Main Results}

\paragraph{\textbf{Zero-Shot Composed Image Retrieval}}

We compare our model in Table~\ref{tab:cir_overall} against various ZS-CIR methods: i-SEARLE~\cite{agnolucci2024isearle}, CIReVL~\cite{karthik2024cirevl},  Pic2Word~\cite{saito2023pic2word}, MLLM-I2W~\cite{bao-etal-2025-mllm}, LinCIR~\cite{gu2024lincir}, CoLLM~\cite{huynh2025collm}, MagicLens~\cite{zhang2024magiclens}, and TransAgg~\cite{liu2023zeroshot}.

Despite being trained on only 5,000 samples from the enhanced LlavaSCo dataset, TransAgg + UNION delivers remarkably competitive performance across all three ZS-CIR benchmarks. On CIRCO, it achieves the best scores in both mAP@10 (34.5) and mAP@50 (38.5), surpassing all other methods, including those trained on datasets with millions of triplets. This result highlights UNION’s effectiveness in improving target representation by aligning better with the fused multimodal query. While it slightly trails in FashionIQ and CIRR benchmarks compared to models trained on much larger datasets (e.g., CoLLM, MagicLens), the performance gap is relatively small (e.g., 31.9 vs. 34.6 on FashionIQ@10), especially considering the drastic reduction in training data size. These findings demonstrate that UNION not only improves semantic alignment and contrastive learning efficiency but also enables high performance under minimal supervision, reinforcing its value for scalable and data-efficient ZS-CIR systems.

\begin{table*}[!h]
    \centering
    \begin{tabular}{|c|c|c|c|c|c|c|c|c|}
        \hline \multirow{2}{*}{\textbf{Method}}& \multirow{2}{*}{\textbf{Backbone}} & \multirow{2}{*}{\textbf{\# Pairs}} & \multicolumn{2}{|c|}{\textbf{Sketchy}} & \multicolumn{2}{|c|}{\textbf{TU-Berlin}} & \multicolumn{2}{|c|}{\textbf{QuickDraw}} \\
        \cline{4-9} & & & \textbf{mAP@200} &  \textbf{Prec@200} & \textbf{mAP} & \textbf{Prec@100} & \textbf{mAP} & \textbf{Prec@200}\\
        \hline DCDL \cite{li2025dcdl} & CLIP-B & 57K/15K/236K & 72.6 & \textcolor{blue}{76.9} & \textcolor{red}{63.4} & \textcolor{red}{74.1} & \textcolor{red}{33.6} & 29.6 \\
        \hline CAT \cite{sain2023cat} & CLIP-B & 57K/15K/236K & 71.3 & 72.5 & \textcolor{blue}{63.1} & 72.2 & 20.2 & 38.8\\
        \hline IVT \cite{zhang2024ivt} & ViT-B & 57K/15K/236K & 61.5 & 69.4 & 55.7 & 62.9 & 32.4 & 16.2\\
        \hline ZSE-SBIR \cite{lin2023zsesbir} & ViT-L & 57K/15K/236K & 52.5 & 62.4 & 54.2 & 65.7 & 14.5 & 21.6\\
        \hline MagicLens \cite{zhang2024magiclens} & CLIP-L & 36.7M & 68.2 & 75.8 & 62.9 & \textcolor{blue}{73.1} & 15.1 & 20.4 \\
        \hline TransAgg & CLIP-L & 5K & \textcolor{blue}{79.6} & 75.8 & 45.4 & 68.2 & 30.1 & \textcolor{red}{43.5}\\
        \hline TransAgg + UNION & CLIP-L & 5K & \textcolor{red}{82.7} & \textcolor{red}{79.9} & 51.0 & 69.8 & \textcolor{blue}{33.4} & \textcolor{blue}{41.5}\\
        \hline
    \end{tabular}
    \caption{Comparison of our method against existing frameworks on three benchmarks of ZS-SBIR task. Except MagicLens and Ours, the others are trained on their own training sets. \textcolor{red}{Red} and \textcolor{blue}{Blue} numbers indicate the best and second-best results.} 
    \label{tab:sbir_overall}
\end{table*}

\paragraph{\textbf{Zero-Shot Sketch-Based Image Retrieval}} 

% We compare our approach with several ZS-SBIR methods sharing the same backbones, including: \textbf{DCDL} \cite{li2025dcdl} presented a dual causal disentanglement framework using two VAEs to separate sketch and image features into exchangeable attributes; \textbf{CAT} \cite{sain2023cat} uses prompt learning and CLIP for performance improvement by specific prompts for sketch, and patch shuffling methods; \textbf{IVT} \cite{zhang2024ivt} solve domain and semantic discrepancies by adopting Vision Transformer for featuring a feature picker and parallel feature adapter; \textbf{ZSE-SBIR} \cite{lin2023zsesbir} proposes a unified, explainable transformer-based model that matches sketches and photos via local patch correspondences, achieving strong zero-shot retrieval without external semantic supervision; and \textbf{MagicLens} \cite{zhang2024magiclens} is trained on their 36.7M-triplet own dataset instead of the benchmark training set.

We compare our approach with several ZS-SBIR methods using the same backbones: 
DCDL~\cite{li2025dcdl}, CAT~\cite{sain2023cat}, IVT~\cite{zhang2024ivt}, ZSE-SBIR~\cite{lin2023zsesbir}, and MagicLens~\cite{zhang2024magiclens}.

Table~\ref{tab:sbir_overall} compares the performance of our UNION-enhanced TransAgg framework against state-of-the-art ZS-SBIR methods across Sketchy \cite{yelamarthi2018zssbir}, TU-Berlin \cite{zhang2016tuberlin}, and QuickDraw \cite{dey2019quickdraw}. With only 5,000 training pairs, TransAgg + UNION achieves the best results on Sketchy (82.7 mAP@200 and 79.9 Prec@200), improving upon the original TransAgg by +3.1 mAP and +4.1 Prec@200. This improvement is particularly notable given that prior methods like DCDL and CAT rely on up to 57K training pairs yet still underperform.

On QuickDraw, UNION reaches the second-best mAP (33.4), close to DCDL (33.6) while providing a substantial +8.0 Prec@200 gain over TransAgg (41.5 vs. 33.5). Similarly, on TU-Berlin, UNION achieves 51.0 mAP, outperforming most baselines and showing a +6.0 mAP improvement over TransAgg, though slightly below DCDL, which benefits from larger-scale training (57K–236K pairs) and task-specific optimisation.

Overall, these results confirm that UNION features effectively enrich target representations and narrow the modality gap between sketches and real images. The consistent improvements in both precision and mAP demonstrate that UNION not only enhances overall ranking quality but also improves top-k retrieval accuracy. This makes it a robust and resource-friendly solution for ZS-SBIR, especially under limited supervision scenarios.

\subsection{Qualitative Results for ZS-IGROT} 

\begin{figure*}
    \centering
    \includegraphics[width=0.7\linewidth]{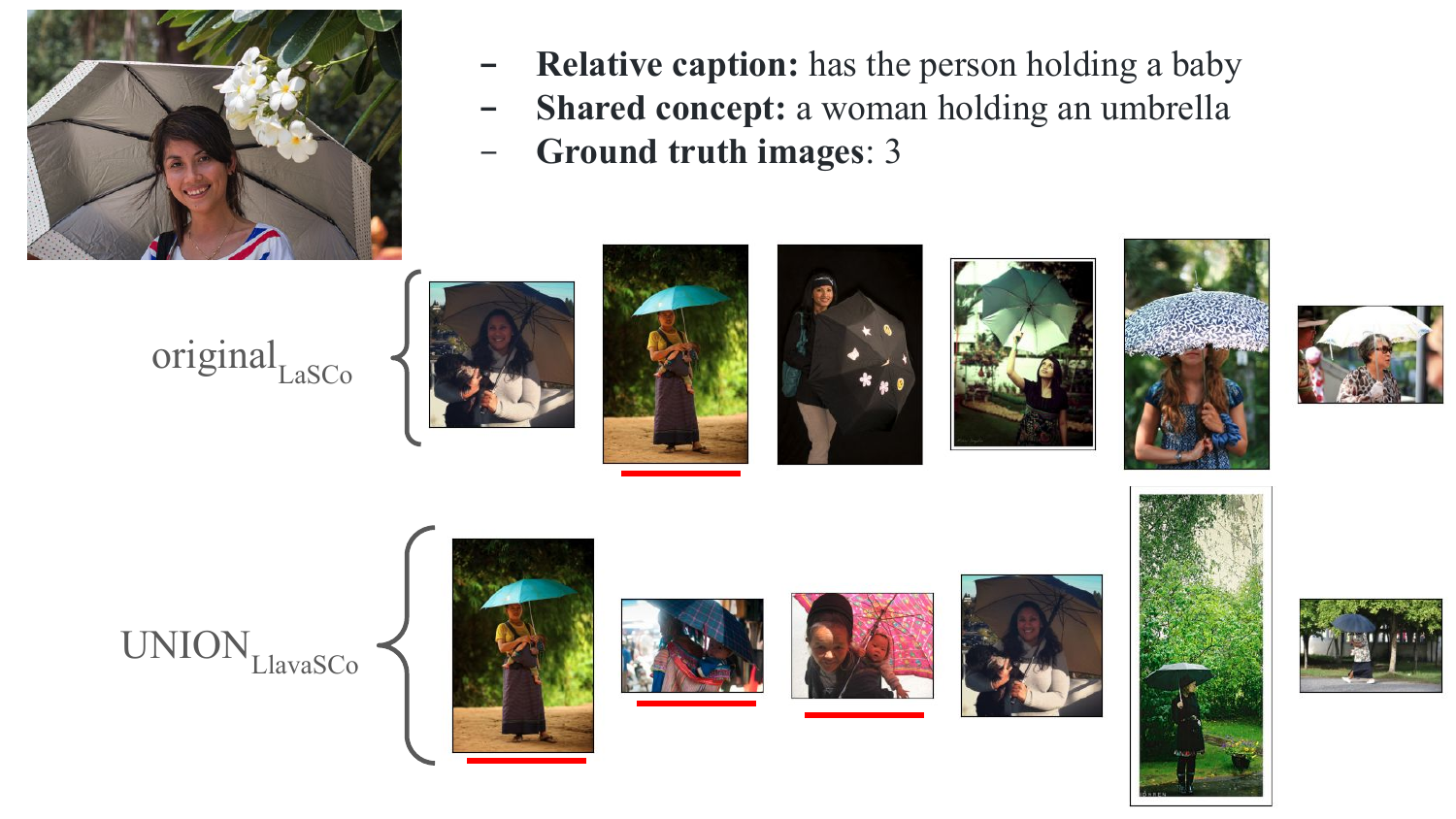}
    \includegraphics[width=0.7\linewidth]{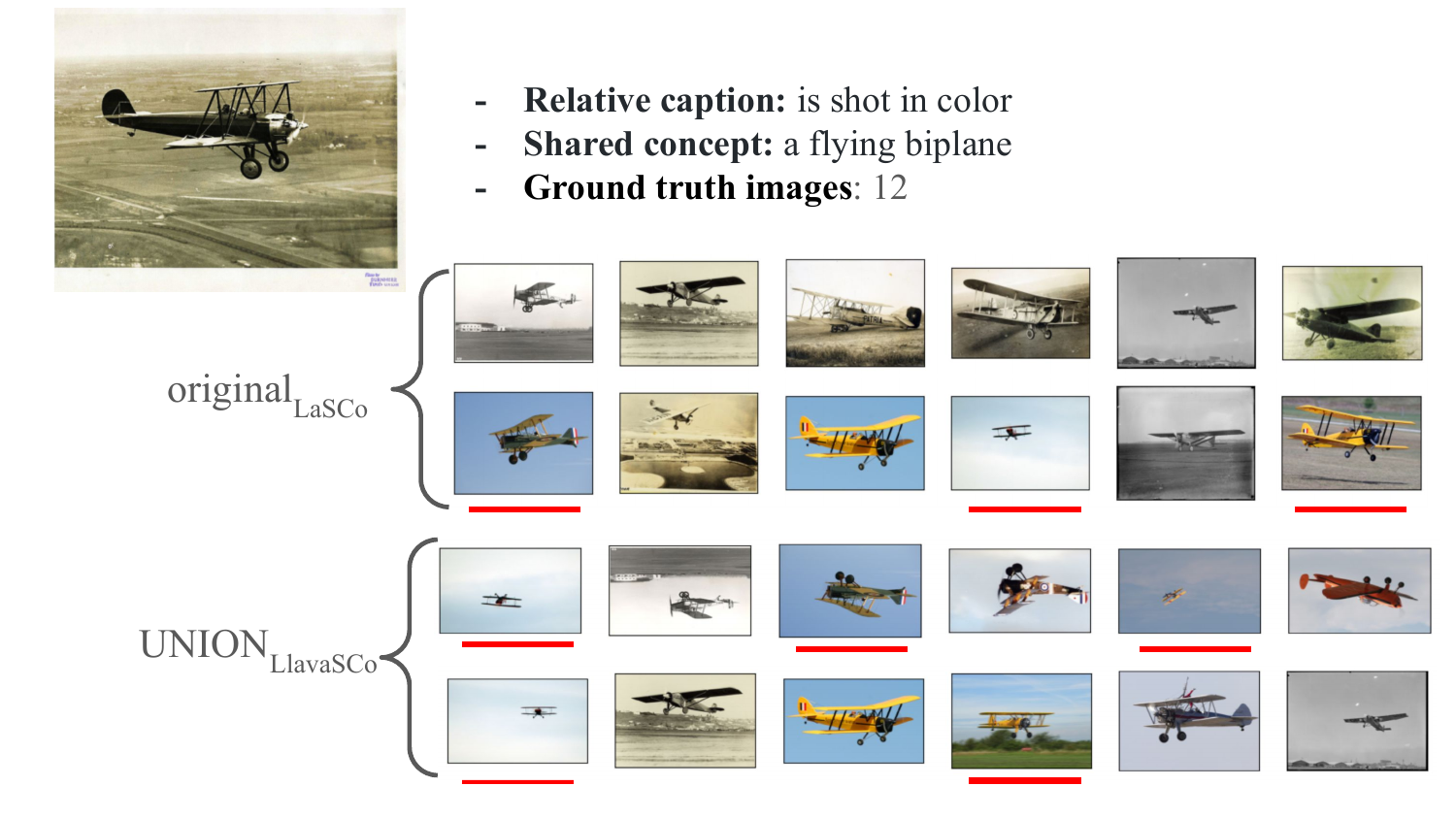}
    \caption{Qualitative Results on CIRCO validation set . The ground truth images are red-underlined.}
    \label{fig:qualitative}
\end{figure*}

Figure~\ref{fig:qualitative} presents qualitative retrieval examples from the CIRCO validation set under the ZS-CIR setting, where evaluation is performed using Recall@$K$, with $K$ corresponding to the number of ground-truth target images per query. We observe that the model enhanced with the UNION feature retrieves target images that more accurately reflect the intended textual modifications compared to the baseline. In cases involving subtle or attribute-level changes—such as alterations in object color, quantity, or context—UNION facilitates more precise semantic alignment. In contrast, the original fixed feature often yields visually similar but semantically mismatched results. These qualitative results further support our claim that UNION enhances compositional reasoning and improves fine-grained retrieval, even in limited supervision scenarios.

\subsection{Ablation Study}

\begin{figure}[!ht]
    \centering
    \includegraphics[width=\linewidth]{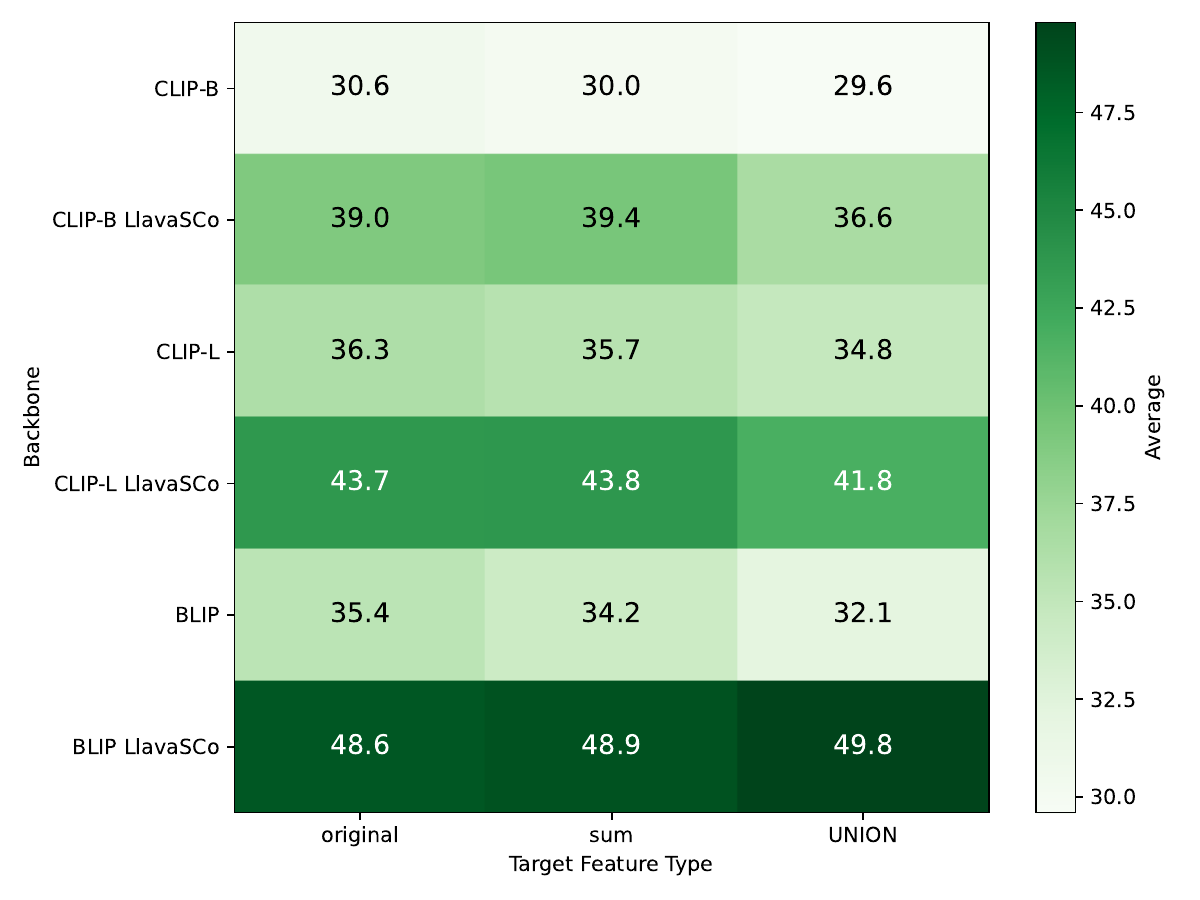}
    \caption{Heatmap showing average performance across different target feature types (original, sum, UNION) and backbones (CLIP-B, CLIP-L, BLIP), with and without enhanced captions from LlavaSCo, on the ZS-CIR task. UNION consistently improves performance in caption-rich scenarios, especially with stronger backbones like BLIP and CLIP-L, while the benefit diminishes slightly when trained without text.}
    \label{fig:cir_heatmap}
\end{figure}

\paragraph{\textbf{Effect of Target Representations and Captions in ZS-CIR}}
Figure \ref{fig:cir_heatmap} presents a heatmap comparing average ZS-CIR performance (the average of all the metrics used ing Table \ref{tab:cir_overall}) across backbones (CLIP-B, CLIP-L, BLIP) and target feature types (original, sum, UNION), with and without enhanced captions from LlavaSCo. Firstly, incorporating LlavaSCo captions consistently boosts retrieval performance for all backbones—highlighting the importance of high-quality, detailed textual supervision in learning fine-grained visual transformations. For example, BLIP’s average score increases from 35.4 to 48.6 (original), and from 32.1 to 49.8 (UNION).

Secondly, UNION shows clear advantages when combined with strong vision-language models and semantically enriched training data. BLIP with LlavaSCo achieves the best overall score (49.8), indicating that UNION benefits from both expressive backbones and rich caption context. However, when trained without captions, UNION may underperform slightly, e.g., on CLIP-B, it trails both original and sum features, suggesting its reliance on latent multimodal grounding even when explicit text is absent.

The results validate the design of UNION as a semantically adaptive representation, especially effective under caption-rich scenarios, while also demonstrating the synergy between improved data quality and retrieval-aware target representations.

\paragraph{\textbf{Evaluating UNION Feature Effectiveness in Zero-Shot Sketch-Based Image Retrieval}}

\begin{figure}[!h]
    \centering
    \includegraphics[width=0.8\linewidth]{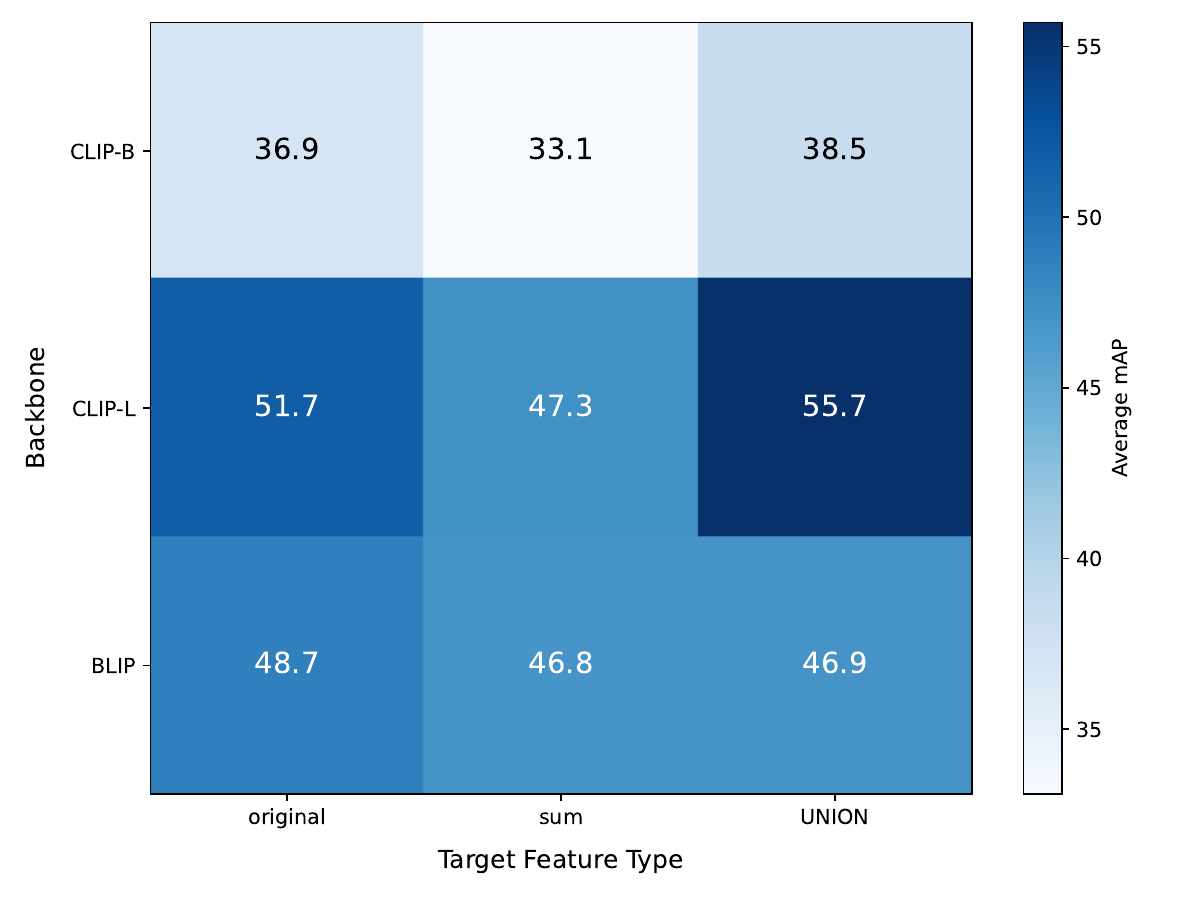}
    \caption{Heatmap of average mAP performance on the ZS-SBIR task for three target feature types across CLIP-B, CLIP-L, and BLIP backbones. The UNION feature outperforms the sum variant and often surpasses the original, particularly with CLIP-L (48.6) and BLIP (44.4), demonstrating its effectiveness in bridging modality gaps between sketches and real images.}
    \label{fig:sbir_heatmap}
\end{figure}

Figure~\ref{fig:sbir_heatmap} presents an updated evaluation of three target feature types (original, sum and UNION) across the CLIP-B, CLIP-L, and BLIP backbones in the ZS-SBIR task, measured by average mAP. Across all backbones, UNION demonstrates superior or comparable performance, showcasing its effectiveness in learning robust retrieval representations with minimal data.

The most notable improvement is seen with CLIP-L, where UNION achieves the highest average mAP score of 55.7, significantly outperforming both the original (51.7) and the sum (47.3) variants. This indicates that UNION is especially beneficial when paired with a strong and expressive vision-language model, helping the network better align sketch queries with photo targets. The improvement margin here suggests that UNION’s feature fusion better captures semantic relationships between modalities, which is a key challenge in ZS-SBIR.

In the CLIP-B setting, UNION also leads with 38.5, outperforming original (36.9) and sum (33.1), highlighting its ability to compensate for limited representational power in smaller models. Interestingly, in the BLIP setting, while all three variants yield closely matched results, UNION still edges out the others slightly (46.9 vs. 48.7 original and 46.8 sum), showing that its benefit persists even in high-capacity backbones but with diminishing marginal returns.

These findings confirm that UNION enhances target representation quality and retrieval accuracy in ZS-SBIR, particularly for mid-sized models like CLIP-L, where the balance between semantic flexibility and model capacity is most advantageous.

\begin{figure}
    \centering
    \includegraphics[width=\linewidth]{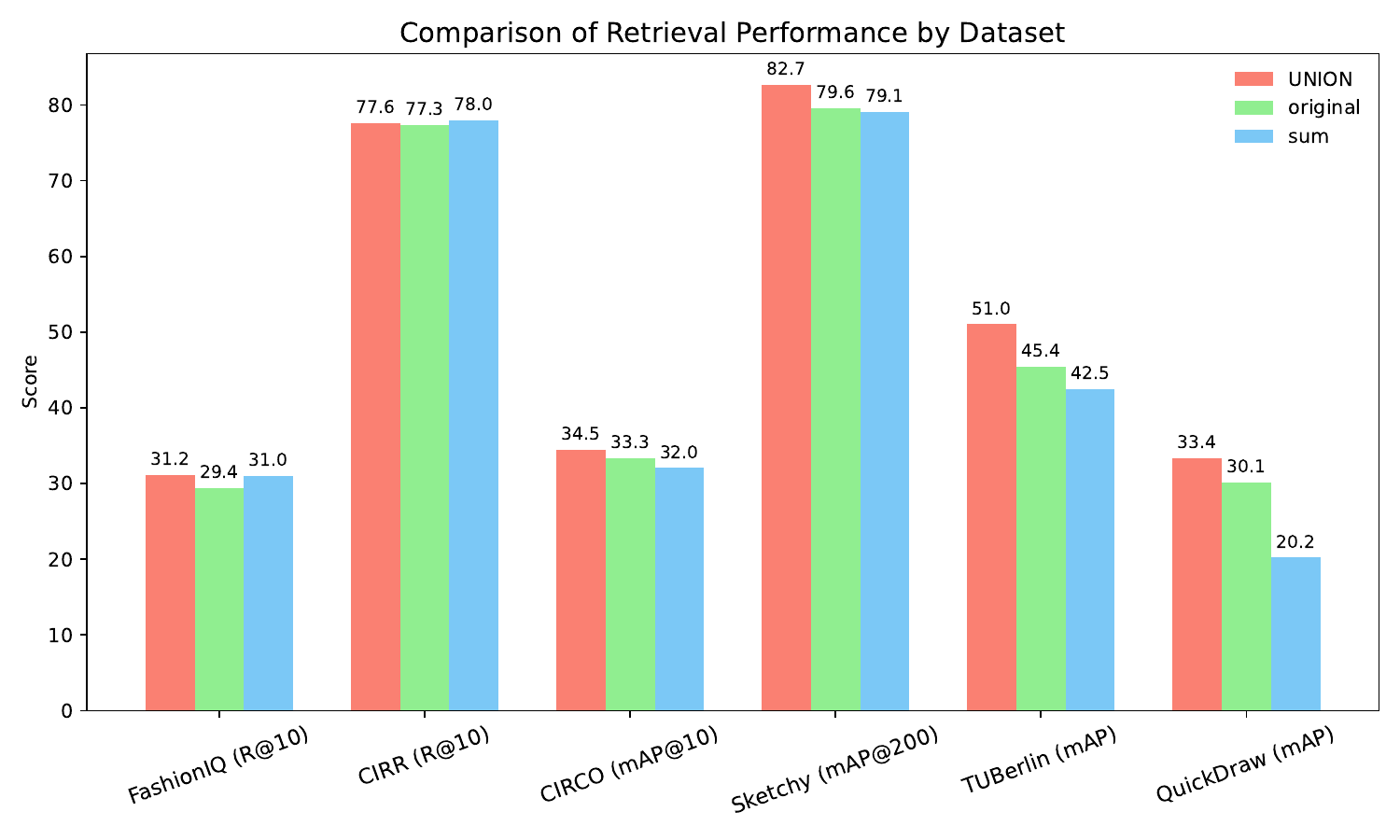}
    \caption{Comparison between three types of target feature: UNION $\mathcal{U}$, original $\mathbf{e}_{t}$, sum $\mathbf{e}_{t}+\mathbf{e}_\eta$. We train on LlavaSCo for ZS-CIR task and Training-Sketchy for ZS-SBIR task.}
    \label{fig:union_comparison}
\end{figure}

\paragraph{\textbf{Comparative Analysis of Target Feature Types}}
Figure \ref{fig:union_comparison} compares the performance of three target feature types—original, sum, and UNION—across six datasets spanning both ZS-CIR (FashionIQ, CIRR, CIRCO) and ZS-SBIR (Sketchy, TU-Berlin, QuickDraw). UNION consistently delivers superior or competitive results, highlighting its robustness across modalities and tasks. In ZS-SBIR benchmarks such as Sketchy and TU-Berlin, UNION clearly leads, achieving the highest scores (82.7 and 51.0, respectively), which indicates its strength in handling sketch-image modality gaps through adaptive semantic alignment.

In the CIR domain, UNION matches or slightly surpasses the original in CIRR (78.0 vs. 77.3) and CIRCO (34.5 vs. 33.3), while staying competitive on FashionIQ. Notably, the sum-based feature consistently underperforms, especially on QuickDraw (20.2 vs. 33.4 with UNION), suggesting that simple additive fusion fails to capture meaningful compositional nuances. These results reinforce that UNION is not only effective in enriching the target representation with latent context, but also generalises well across structured and unstructured retrieval tasks without requiring architectural changes.

\paragraph{\textbf{Limitations}}  
While our approach demonstrates strong performance with limited supervision, it also presents some practical limitations. Firstly, the construction of the LlavaSCo dataset requires running LLaVA-1.6 Mistral to generate captions for 360k reference-target image pairs, which is computationally expensive and time-consuming at scale. Secondly, the inference stage incurs additional latency due to feature processing: generating target embeddings with CLIP-B, CLIP-L, and BLIP across original, sum, and UNION settings takes on average 271 seconds, 334 seconds, and 656 seconds respectively. This highlights a trade-off between representation quality and computational efficiency, which could be further addressed in future work through faster captioning or feature caching strategies. 

\section{Conclusion}

In this work, we explore the challenge of \textbf{Image-Guided Retrieval with Optional Text} under limited supervision. We introduce \textbf{LlavaSCo}, a caption-enhanced dataset constructed from LaSCo \cite{Levy2024lasco} using LLaVA-generated descriptions, and show that training on just 5,000 samples from \textbf{Training-Sketchy} is sufficient to achieve strong retrieval performance on ZS-SBIR task. Central to our approach is the proposed \textbf{UNION} feature, which replaces traditional fixed target embeddings by combining image features with a null-text prompt. This representation improves semantic alignment with the query and enhances contrastive discrimination, all without modifying pretrained vision-language backbones. However, UNION relies on the presence—or generation—of informative captions to fully leverage linguistic cues, and it introduces additional inference overhead when embedding large image pools. Future directions include exploring more efficient captioning strategies and optimising inference time.

\section*{Appendix A: LlavaSCo Caption Generation.} \label{appendix_a:llavasco}

We observe that LaSCo triplets often contain weak or generic relational captions that do not clearly convey the transformation from reference to target image. To strengthen this link, we enhance each triplet using LLaVA~\cite{liu2023improvedllava}, a multimodal model trained to generate fine-grained image descriptions.

For each target image, we generate a caption using the following prompt:

\begin{quote}
    \textit{Describe the image in one sentence with details.}
\end{quote}

We extract only the first sentence to ensure brevity and consistency. The new relative caption is formed as:

\[
\text{RelCap}_{\text{new}} = \text{RelCap}_{\text{old}} \text{ with } \text{GenCap}_{\text{target}}.
\]

From the full LaSCo dataset (360k triplets), we select a subset of 5,000 refined triplets to form the training set for our CIR experiments. The original validation and test sets are kept unchanged. This refinement helps bridge the modality gap for our UNION representation by improving the textual grounding of each training instance.

\section*{Appendix B: Additional Evaluation Metrics.}\label{appendix_b:metrics}

Besides the standard retrieval metrics, we introduce two complementary evaluation criteria to better assess the effectiveness of UNION: \textbf{Median Rank} (MdR) and \textbf{Mean Average Precision per Ground Truth Number} (mAP/GTN).

First, Median Rank (MdR) measures the median position of the ground truth image in the ranked retrieval list across all queries; a lower MdR indicates better retrieval precision. Second, we observe that as the retrieval cutoff $k$ increases, Precision scores typically decline. To mitigate this bias, we adopt mAP/GTN, which computes mean average precision where the number of retrieved candidates is set exactly to the number of ground truth images for each query—providing a fairer evaluation of methods in multi-ground-truth settings.

We evaluate these two metrics on the CIRCO validation split and the Sketchy dataset, both of which provide multiple correct target images per query.

\begin{table}[!h]
    \centering
    \begin{tabular}{|c|c|c|c|}
        \hline \textbf{Dataset} & \textbf{Method} & \textbf{MdR} $\downarrow$ & \textbf{mAP/GTN} $\uparrow$ \\
        \hline \multirow{3}{*}{\textbf{CIRCO}$_\text{val}$} &  original$_\text{LaSCo}$ & 6.9 & 24.9 \\
         & original$_\text{LlavaSCo}$ & 5.8 & \textcolor{red}{32.7} \\
         & UNION$_\text{LlavaSCo}$ & \textcolor{red}{5.3} & 32.2 \\
        \hline \multirow{3}{*}{\textbf{Sketchy}} & original & 6.7 & 59.9 \\
         & sum & 6.5 & 57.1 \\
         & UNION & \textcolor{red}{6.1} & \textcolor{red}{68.8} \\
        \hline
    \end{tabular}
    \caption{Performance comparison under Median Rank (MdR) and Mean Average Precision per Ground Truth Number (mAP/GTN) on CIRCO\textsubscript{val} and Sketchy datasets. UNION consistently improves retrieval ranking and robustness across multiple ground-truth settings.}
    \label{tab:new_metric}
\end{table}

As shown in Table~\ref{tab:new_metric}, UNION consistently outperforms baseline target features on both CIRCO and Sketchy datasets under the new metrics. On CIRCO\textsubscript{val}, UNION achieves the lowest MdR (5.3) and a highly competitive mAP/GTN (32.2), confirming its ability to retrieve relevant targets with fewer distractors. On Sketchy, UNION yields a substantial improvement, boosting mAP/GTN to 68.8, compared to 59.9 (original) and 57.1 (sum), while also lowering the MdR to 6.1. These results further demonstrate that UNION enhances both ranking quality and retrieval robustness, especially when multiple correct answers exist.

\section*{Acknowledgment}

    This publication has emanated from research supported in part by research grants from Science Foundation Ireland (SFI) under grant numbers  SFI/13/RC/2106\_P2 and 18/CRT/6223, and co-funded by the European Regional Development Fund.

\bibliographystyle{IEEEtran}
\bibliography{main_ICDM}

\end{document}